\documentclass[8pt,twocolumn,preprintnumbers,amsmath,amssymb,nofootinbib,superscriptaddress,prl]{revtex4-2}

\usepackage{amsmath,amssymb,graphicx} 

\usepackage{siunitx}
\usepackage[usenames,dvipsnames]{color}

\newcommand{\mc}[1]{{\color{black} #1}}

\graphicspath{{figures/}}

\begin{document}

\title{Creation of a black hole bomb instability in an electromagnetic system} 

\author{M. Cromb}
\affiliation{School of Physics and Astronomy, University of Southampton, SO17 1BJ, Southampton, UK.}
\author{M.C. Braidotti}
\affiliation{School of Physics \& Astronomy, University of Glasgow, G12 8QQ Glasgow, UK}
\author{A.~Vinante}
\affiliation{Istituto di Fotonica e Nanotecnologie - CNR and Fondazione Bruno Kessler, I-38123 Povo, Trento, Italy}
\author{D. Faccio}
\affiliation{School of Physics \& Astronomy, University of Glasgow, G12 8QQ Glasgow, UK}
\author{H. Ulbricht}
\email[Correspondence email address: ]{H.Ulbricht@soton.ac.uk}
\affiliation{School of Physics and Astronomy, University of Southampton, SO17 1BJ, Southampton, UK.}


\begin{abstract}
{The amplification and generation of electromagnetic radiation by a rotating metallic or lossy cylinder, first theorized by Zel'dovich in the 1970's, is tightly connected to the concepts of quantum friction, energy extraction from rotating black holes and runaway mechanisms such as black hole bombs. Despite recent advances including acoustic analogues of the Zel'dovich effect and the observation of a negative resistance in a low-frequency electromagnetic model, actual positive signal amplitude gain, the spontaneous generation of electromagnetic waves and runaway amplification effects have never been experimentally verified.
Here, we demonstrate experimentally that a mechanically rotating metallic cylinder not only definitively acts as an amplifier of a rotating electromagnetic field mode but also, when paired with a low-loss resonator, becomes unstable and acts 
as a generator, seeded only by noise. The system exhibits an exponential runaway amplification of spontaneously generated electromagnetic modes thus demonstrating the electromagnetic analogue of Press and Teukolsky's `black hole bomb'. The  exponential amplification from noise supports theoretical investigations into black hole instabilities and is promising for the development of future experiments to observe quantum friction in the form of the Zel'dovich effect seeded by the quantum vacuum. 

}

\end{abstract}

\maketitle

{\bf{Introduction.}}
In 1971 Yakov Zel’dovich predicted that an absorbing axially-symmetric body rotating at rotational frequency $F$ and scattering incident waves of angular momentum order $m$ could somewhat counter-intuitively amplify those waves if their frequency $f$ satisfies \cite{zeldovichGeneration1971}
\begin{equation} \label{e:zeldy}
    f<mF.
\end{equation}
This condition is equivalent to the rotational Doppler-shifted co-rotating mode frequency ($f_- = f -  mF$) becoming negative in the body's rotating frame. While Zel'dovich considered the case of a metal cylinder scattering electromagnetic waves \cite{zeldovichAmplification1972,zeldovichRotating1986}, he emphasised that this  amplification (also known as rotational superradiance \cite{bekensteinMany1998}) is a general effect rooted in thermodynamics and should therefore hold true for any rotating absorber.\\
Indeed, his prediction was directly inspired by the Penrose process, a means of extracting energy from the ultimate absorber: a rotating black hole. The rotating spacetime creates an ergoregion around the black hole horizon where matter and waves can have negative energy. Penrose envisioned that an object scattering with this ergoregion, could split into two and lose a negative energy component into the black hole, while the positive energy part escapes having gained energy from the black hole rotation \cite{penroseGravitational1969,penroseExtraction1971}. This Penrose superradiance has been tested recently in analogue systems in the laboratory \cite{torresRotational2017,braidottiPen2020,braidottiMeasurement2022} and has been proposed as part of a mechanism producing relativistic jets of quasars~\cite{williamsExtracting1995}. \\
This link between the Zel'dovich effect and black hole thermodynamics holds also in the quantum realm. Zel'dovich predicted that a rotating absorber could spontaneously amplify electromagnetic (EM) fields out of the quantum vacuum \cite{zeldovichRotating1986}, ceding its rotational energy and slowing down. This implication directly inspired \cite{thorneBlack1994} Hawking’s famous prediction that even \textit{without} rotation, \textit{any} black hole should slowly radiate its energy away \cite{hawkingBlack1974,hawkingParticle1975}. However, this quantum vacuum rotational amplification is very weak and hence difficult to observe. For this reason, Zel'dovich speculated that forming a low-loss resonator by encircling the cylinder with a mirror could amplify this very weak signal \cite{zeldovichGeneration1971,zeldovichAmplification1972}. This generation mechanism was detailed further in 1972 with Press and Teukolsky's 'black-hole bomb' concept \cite{pressFloating1972,cardosoBlackhole2004}. 
A rotating black hole scatters and superradiantly amplifies the impinging modes that satisfy Eq.~(\ref{e:zeldy}). Surrounding the black hole with a mirror will reflect scattered modes back towards the hole to be re-amplified. If the mirror is sufficiently reflective, the energy lost at the mirror can be smaller than the energy gained from the black hole.  With this positive feedback, such amplified signals grow exponentially, and the system becomes unstable to any random noise seed. The field energy trapped by the mirror grows until it is either released through a controlled opening (another proposed power source), or if unchecked, until the mirror can no longer take the pressure and explodes. A third mechanism is also possible: Cardoso et al. showed that the exponential amplification can also be switched off, if the black hole loses too much angular momentum before the mirror explodes, at which point the system is no longer unstable and the condition in Eq.~(\ref{e:zeldy}) is no longer satisfied~\cite{cardosoBlackhole2004}. The same instability conditions and behaviors can also occur with Zel'dovich's electromagnetic cylinder case \cite{cardosoBlackhole2004}.   \\
\begin{figure*}[t]
    \centering
    \includegraphics[width=1\textwidth]{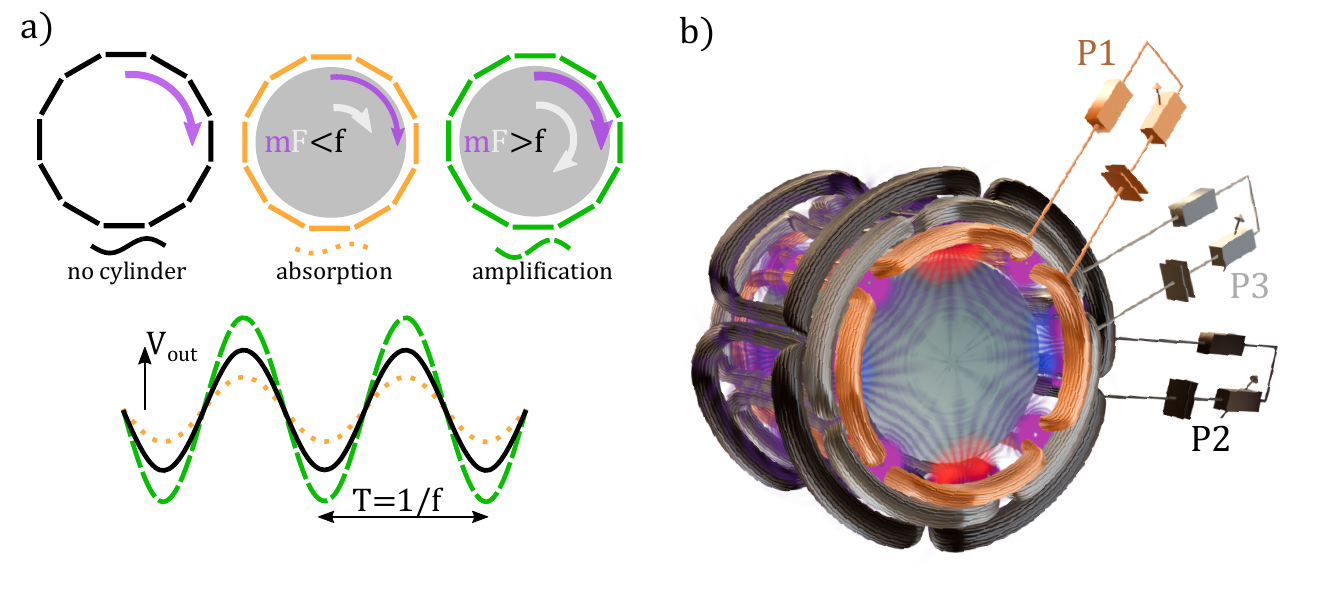}
    \caption{{\bf{Concept and schematic layout of experiment.}}a)  Diagram showing the Zel'dovich amplification condition. For a given input field, the measured output amplitude depends on the presence of the (absorbing) metal cylinder and its rotation speed: with no cylinder, there is no effect; with the internal cylinder in place and rotating slower than the rotating EM field, absorption is increased; if the cylinder rotates faster than the EM field, amplification occurs.  b) {Schematic overview of the full experiment. Three sets of external coils surround the internal aluminium cylinder. Each coil is driven by an RLC circuit (P1, P2, P3) with a variable resistor that is used to tune the losses. We plot the numerically calculated magnetic field lines (at a given instant) in purple on the cylinder. Red and blue areas indicate North and South magnetic poles, respectively. }}
    \label{fig:StatorField}
\end{figure*}
Amplification from a rotating absorber was successfully demonstrated for acoustic waves~\cite{crombAmplification2020,liuExperimental2022a}. For the case of EM waves, notwithstanding the significantly more prohibitive experimental conditions, a recent work measured Zel'dovich amplification by showing that the rotation of a metallic cylinder induces a negative resistance in an electromagnetic circuit (indicating that amplification is occurring, even if losses still dominate the overall behaviour)~\cite{braidottiAmplification2024}.  \\
In this work, we present an experimental study that relies on a rotating magnetic field generated by a 3-phase stator with an internal spinning metallic cylinder. The 3-phase arrangement allows us to generate a magnetic field on the internal cylinder that has a definite rotational direction. At the same time, the external circuit with the stator also acts as a reflector. Thus, the system satisfies the experimental conditions speculated by Zel'dovich for the observation of spontaneous generation and also the conditions outlined by Press et al. for black hole bombs.  Indeed, the experimental conditions implemented show net amplification, despite the internal losses of the circuitry. We also observe the spontaneous generation of waves, seeded only by background noise. This generation exhibits a runaway exponential growth, also known as self-oscillation, of the EM waves in analogy to a black hole bomb. Finally, by modifying the operating conditions of the rotating cylinder, we also observe the regime, described by Cardoso et al. \cite{cardosoBlackhole2004}, where this instability and the exponential amplification switches off due to loss of rotational energy in the cylinder. \\
{\bf{Experimental Setup.}}
Figure~\ref{fig:StatorField}a) shows a visualisation of the Zel'dovich amplification condition, Eq.~(\ref{e:zeldy}). There is: no effect in the absence of the cylinder; an increased absorption if the cylinder rotates slower than the angular phase velocity of the rotating field mode, $f/m$ (purple arrow); amplification occurs only if the cylinder is rotating in the same direction and faster than $f/m$. These 3 conditions are schematically shown in Fig.~\ref{fig:StatorField}a) as 3 different oscillating voltage amplitudes that we measure from our circuit. 

\begin{figure*}[t]
    \includegraphics[width=0.48\linewidth]{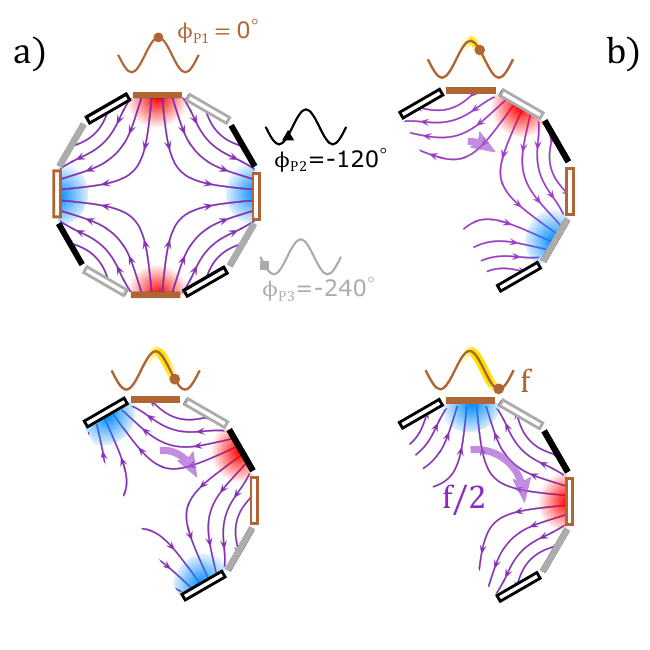}
    \includegraphics[width=0.48\linewidth]{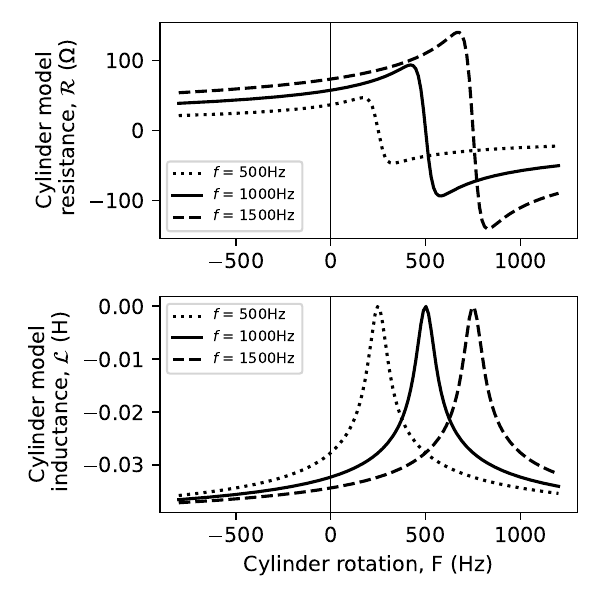}
    \caption{{\bf{Rotating magnetic field and amplification conditions.}} a) Numerically simulated field lines inside the coils for 4 different phase values of the P1 current (P2 and P3 are retarded by -120 and -240 deg, respectively). The North pole (indicated as a red shaded area) at the top coil rotates by 1/4 of a cycle when the circuit currents vary by 1/2 a cycle (i.e. the magnetic field rotates by a full cycle every two current cycles).  b) Theory: the graphs shows the resistance and inductance of the internal cylinder only as a function of the rotation frequency of the internal metal cylinder for three different external quadrupole EM field rotation frequencies. A co-rotating cylinder will lead to a negative cylinder resistance (hence amplification) above the threshold $F = f/2$ and increased resistance (hence increased losses) below this threshold.}
    \label{fig:ZelSimulation}
\end{figure*}
Figure~\ref{fig:StatorField}b) shows a schematic overview of the experimental setup.  
A rotating magnetic field is generated by a three phase induction motor, consisting in a stator with three independent RLC circuits (P1, P2, P3). An aluminium cylinder, spun by a brushless DC motor, is nested inside the stator, with only a small {$\sim1$ mm} air gap separating it from the stator (for details on the experimental apparatus, see Methods).
The magnetic field is rotated in time by ensuring that the three circuits are 120$^{\circ}$ out of phase with respect to each other, and the relative sign of the phase shifts determine the direction in which the magnetic field rotates. The design of the coils determines the shape of the magnetic field lines and leads to a quadrupole rotating mode. This field is numerically simulated and shown as an overlay in purple on the cylinder for a fixed time instant in Figure~\ref{fig:StatorField}b).
The quadrupole rotating mode implies that we have a mode with orbital momentum $m=2$, i.e., the magnetic field completes half a rotation for every $2\pi$ cycle of the sinusoidal current in the circuits (shown in more detail for four different times across a half-cycle of the current in Fig.~\ref{fig:ZelSimulation}a)). \\
In our experiment, each RLC circuit is also used to store electromagnetic energy at its resonant frequency, $f_{res}$. This resonance acts as a lossy mirror (analogue to the black hole bomb mirror) confining the EM radiation around the cylinder, thus increasing the interaction.\\ 
{Figure~\ref{fig:ZelSimulation}b)} shows how the effective resistance and inductance of the cylinder in the  rotating quadrupole  mode ($m=2$) is predicted by our model to change with cylinder rotation rate, $F$ (model details are reported in Methods). When the cylinder is co-rotating with the mode ($F>0$), we see a transition to  negative resistance (i.e. loss transitions to gain)  when $F>f/2$ (we show curves for 3 different values of $f$). The model also predicts that the inductance  changes with rotation and this changes the RLC resonant frequency accordingly. We note here that a negative resistance in the cylinder does not imply a net positive amplitude gain in the full system (e.g. in the circuits P1, P2 and P3) as these will always have an additional resistance and hence a loss term that needs to be overcome by the cylinder amplification. We therefore purposely include a variable resistor $R_{var}$ in each circuit with which we can tune the RLC losses. \\ 
An important detail in the following is the operation regime of the motor driving the internal metallic cylinder. We choose two different measurement settings for the motor drive to observe two distinctly different regimes and better understand the energy exchange dynamics between the mechanically driven internal cylinder and the EM field in the stator: a) a closed-loop setting, where we fix the motor rotation frequency $F$ - this setting will allow us to observe net Zel'dovich amplification and reach the unstable black hole bomb regime; b) an open-loop setting, which allows the motor rotation rate to change - in this case, we will observe wave generation and exponential amplification from noise and then the switching off of the unstable regime due to loss of rotational energy in the cylinder.\\ 
{\bf{Results: Zel'dovich amplification and black hole instability threshold.}}
\begin{figure*}[t]
    \includegraphics[width=0.48\linewidth]{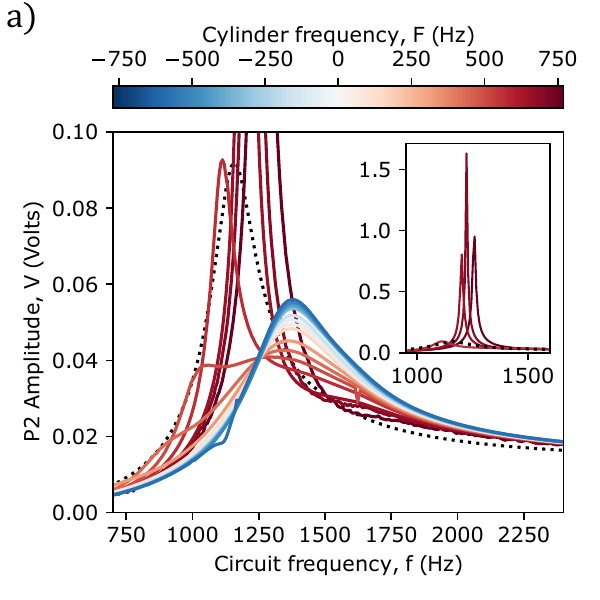}
    \includegraphics[width=0.48\linewidth]{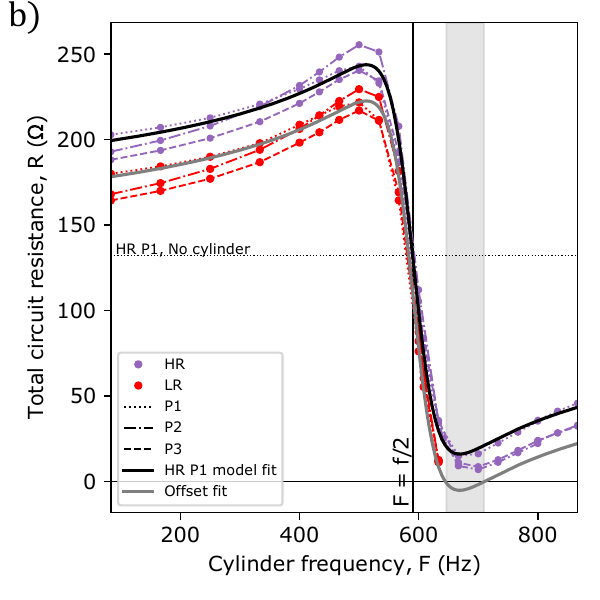} 
    \caption{
    {\bf{Experimental results: closed-loop motor settings.}} a) Shows the measured rms voltage ($V_o$) for increasing EM wave frequency and for varying cylinder rotation rates (blue curves indicate counter-rotating and red curves indicate co-rotating). Also included is the no-cylinder case (black dotted line), which peaks at 0.0916 V.  The inset shows the highest amplitude peaks, which are an order of magnitude above the no-cylinder case. A peak gain factor of 17.6x is observed for the case of the cylinder rotating at 700 Hz.  b) Total resistance $R$ in the {three circuits P1, P2, P3} for a fixed circuit frequency $f=1181$~Hz and different cylinder rotation frequencies $F$. Purple dots are data for the high resistance circuits, red dots for the low resistance circuits. The vertical line indicates the Zel'dovich threshold rotation, the dotted horizontal line indicates the $R_{circ} = 131~\Omega$ measured P1 resistance without the cylinder (for the high resistance case). The numerical model {fits} to the {P1} high resistance data is also plotted (black solid line), with the coupling strength $A$ and the constant resistance present without the cylinder $R_{circ}$ as free fit parameters. We find $R_{circ}=129~\Omega$, which matches the measured value well, and $A=0.397$. The grey line is the fit offset by $\Delta R = -21.2~\Omega$ for P1 ($\Delta R = R_{var}^\text{LR} - R_{var}^\text{HR}$) 
    indicating the expected total resistance in the low resistance case. The shaded region indicates the range of cylinder frequencies where we expect to see a net exponential amplification of a $f=1181$~Hz signal.
    }
    \label{fig:ZelSpeeds}
\end{figure*}
We first set the cylinder motor at a fixed rotation speed in the closed-loop configuration. We choose a relatively high value for the variable resistor $R_{var}\approx 24$ ohms (Table~\ref{tab:VoltResValues150} in Methods) so as to ensure that overall, the total loss (circuit plus cylinder) dominates over the cylinder gain for all conditions. 
Fig.~\ref{fig:ZelSpeeds}a) shows the voltage amplitude in circuit P2 as we vary the EM frequency $f$ in the circuit (similar results are obtained in all three circuits). The various curves are for different rotation frequencies $F$ of the cylinder, as indicated in the legend. A clear resonance is observed in the `no-cylinder' case (black dotted curve) - this resonance arises, as discussed, from the RLC circuit and is determined by the choice of inductance and capacitance values. It is around this resonance that we have energy accumulation in the circuit, akin to energy accumulation in a cavity that enhances the interaction of the EM field with the cylinder. It is around this resonance that we focus our attention.
We see variations in the resonant peak amplitude as a result of the cylinder rotation.
The blue curves in Fig.~\ref{fig:ZelSpeeds}a) indicate the counter-rotating cases, which all exhibit lower maximum amplitudes compared to the no-cylinder case (dotted curve), i.e. the presence of a counter-rotating cylinder increases the total losses. The red curves indicate the co-rotating cases and lead to reduced losses with very significant loss-reduction factors, i.e., $>10 \times$ when Eq.~(\ref{e:zeldy}), $F > f/2$ is satisfied, i.e. for $F\gtrsim600$ Hz (see 3 highest peaks in Fig.~\ref{fig:ZelSpeeds}a) inset). Furthermore, the resonant peak frequency also shifts with the cylinder rotation speed: the resonance frequency is minimum for $|f/2-F|=0$, when the cylinder is  co-rotating with the field - or there is no cylinder (dotted curve) - and increases for $|f/2-F|>0$, due to the additional cylinder inductance, as also discussed in Fig.~\ref{fig:ZelSimulation}b).\\ 
\begin{figure*}[t]
    \centering
    \includegraphics[width=0.96\linewidth]{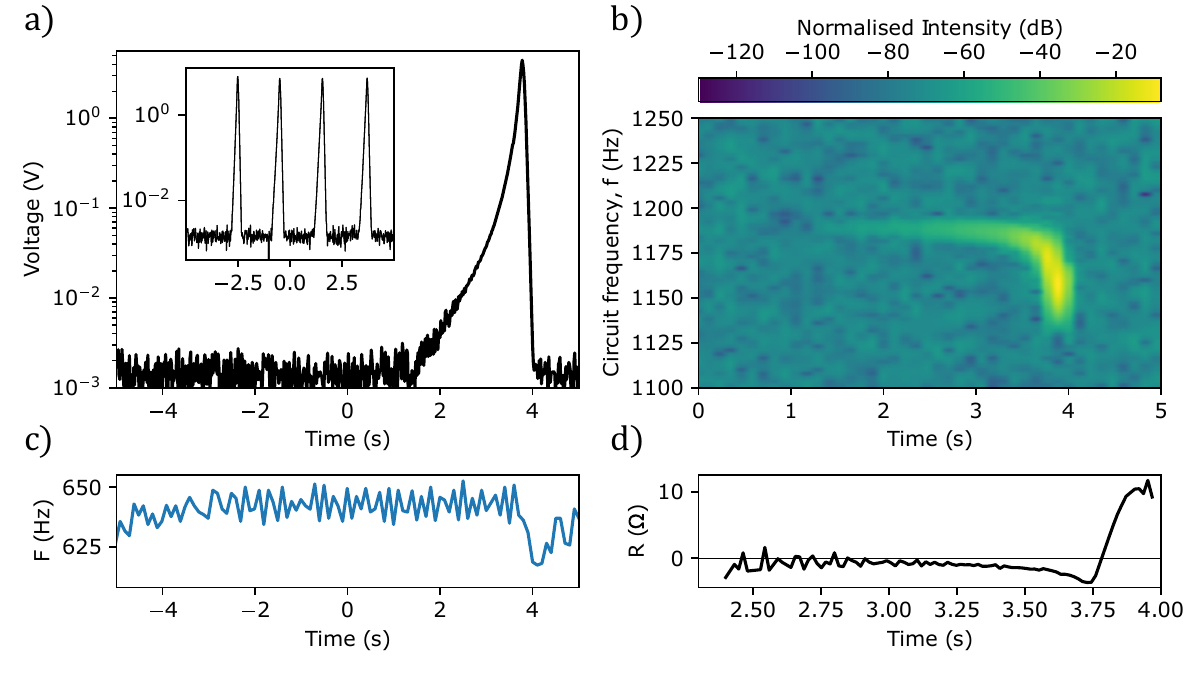}
    \caption{
    {\bf{Experimental results: open-loop configuration.}} a) The increase of the voltage measured over the 5~$\Omega$ resistors in the circuits, on a log scale, when the cylinder is set to rotate at $F= +643$ Hz and driven in the open-loop feedback configuration. No input signal is supplied to the circuits. We show the envelope of the signal measured in time, bandpass filtered to the 1100-1250 Hz band shown in b). The inset shows the circuit evolution over an equal measurement time for the cylinder driven to $F=+660$~Hz, which grows more rapidly, evidencing the cyclical self-halting and amplifying  behaviour of the system. b) Spectrogram of the measured signal (dB taken with reference to max value), highlighting that the signal is seeded by the noise floor in a narrow band of frequencies that correspond to the most negative total resistance. c) Measured cylinder rotation rate $F$ over time, exhibiting a marked decrease in correspondence to the exponential growth of the circuit voltage - this is evidence of the system converting mechanical rotational energy into EM energy. d) Total resistance $R$ extracted from the signal that is negative (in the initial linear region around -0.75 ohms) until the cylinder slows down sufficiently that the cylinder gain (negative resistance) is overcome by the circuit resistance, resulting in a positive total net resistance.}  
    \label{fig:self-osc}
\end{figure*}
\noindent Figure~\ref{fig:ZelSpeeds}b) shows the measured \emph{total} circuit resistance for all three P1, P2, P3 circuits, for varying cylinder rotation rates at a fixed circuit EM frequency of 1181 Hz. The solid black curve indicates the numerical model with $R_{circ}=129$ ohms ($R_{circ}$ is the circuit resistance without the cylinder). The model fits the data well and the positive resistance values confirm that overall, the total system (cylinder plus circuit) is not amplifying. We note that the total resistance approaches zero around $F=680$ Hz, indicating a nearly perfect cancellation of the circuit positive resistance with the cylinder negative resistance. By decreasing the value of the variable resistors $R_{var}$, we expect these curves to shift downwards. In this way, we can push the curve around $F=680$ Hz into a region of total negative resistance, i.e., reaching absolute gain. These measurements are also shown for all three circuits (red curves), overlaid with the numerical model (gray curve). As can be seen, the measurements do not actually continue into the negative resistance region: any attempts to perform measurements at cylinder rotations $F>640$ Hz led to a runaway amplification causing the circuit resistor to explode, a very eloquent hint that the `black hole bomb' regime has taken over.\\
The runaway behaviour and explosions are a direct result of driving the motor with fixed rotation speed in the closed-loop configuration, i.e. when the rotational energy is extracted from the cylinder, the feedback loop feeds new power into the motor to keep it at the same speed. Thus, the system can continuously draw increasing amounts of power from the cylinder motor to feed the increasing voltage and current in the circuits.\\
{\bf{Results: Noise amplification and self-limitation of the instability.}}
The exponential trend of the instability can be measured by changing the measurement settings to the open-loop configuration, while keeping $R_{var}$ at the low value to have a net total gain. 
The circuit  can now only feed off the rotational energy of the cylinder for a limited time before slowing it down - the cylinder rotation rate will then drop below the instability condition and the total resistance will return to positive values, switching off the instability before the resistor explodes. Furthermore, we remove the circuit input signal, leaving only noise as a seed. 
The measurements in this regime are shown in Fig.~\ref{fig:self-osc}. We set the cylinder to rotate at $~643$Hz such that we are close to the `negative frequency' region for our circuit (shaded area in Fig.~\ref{fig:ZelSpeeds}b). Figure~\ref{fig:self-osc}a) shows the time trace of the voltage in circuit P2: initially, we observe only noise while the cylinder is rotating at a constant speed (see Fig.~\ref{fig:self-osc}c)). A growing signal, initially masked by the detection noise. appears after a few seconds as a result of amplification of the circuit noise. \mc{We measured the phase of the spontaneously generated field and find that it co-rotates with the cylinder, as predicted for a signal generated through this `black hole bomb' instability (see SM).} 
The inset in Fig. \ref{fig:self-osc}a) shows the circuit dynamics for a higher cylinder rotation ($660$ Hz) and so greater number of e-folds, where we see the `oscillation' between stable and unstable dynamics due to the motor periodically slowing down as it loses mechanical energy to the circuit and then speeds up again as it falls below the exponential amplification frequency range. This regime represents the Cardoso et al. prediction for black hole bomb instability for the Zel'dovich cylinder case \cite{cardosoBlackhole2004}.\\
We also see,  Fig. \ref{fig:self-osc}b) that the amplified signal shifts to lower frequency values, before the amplification switches off. Indeed a decrease in cylinder rotation speed causes the resonance of the circuit to shift toward lower frequencies (see Fig.~\ref{fig:2DAmpPhase_HR}, Methods). This process continues until the cylinder speed becomes too low, exits the instability condition (the total negative resistance region) and the signal dissipates away. 
Another interesting feature in Fig.~\ref{fig:self-osc}a) is the observation of an initial exponential growth rate that at later times becomes super-exponential. This super-exponential behaviour corresponds to a nonlinear decrease of the total resistance $R$ in Fig.~\ref{fig:self-osc}d) and signals a clear departure from the standard black hole bomb theory \cite{zeldovichGeneration1971,zeldovichAmplification1972,pressFloating1972,cardosoBlackhole2004}. These nonlinearities of the system will be investigated in future research. 

{\bf{Discussion and conclusions.}}
The black hole laser \cite{jacobson1999}, predicted by Jacobson et al. as a way to amplify analogue Hawking radiation, and Press and Teukolsky's black hole bomb \cite{pressFloating1972} are examples of black hole instabilities that are enhanced by the presence of a surrounding `mirror' that increases the mode density in the vicinity of the unstable region. The experiments presented here are a direct realisation of the rotating absorber amplifier first proposed by Zel'dovich in 1971 and later developed by Press and Teukolsky into the concept of black hole bomb. In both these cases the amplifier is seeded by the quantum vacuum and this amplification of quantum fluctuations from a rotating absorber still remains to be observed experimentally. In this work, the amplifier is operated at room temperature and is therefore seeded by noise that dominates vacuum fluctuations by many orders of magnitude. Nevertheless, the physical ingredients are as proposed more than 50 years ago. The results show that extraction of rotational energy from an absorber with exponential amplification of EM waves can be observed at low-frequencies, where the conditions for negative energies (or negative resistances) can be met. Furthermore, it also shows how this unstable regime can be switched on and off as predicted for the black hole bomb \cite{cardosoBlackhole2004}. A challenge for the future remains the observation of spontaneous EM wave generation and runaway amplification seeded from the vacuum. However, based on the results presented here, this now remains a purely technological (even if very hard) feat. As pointed out by Unruh, any quantum noise amplifier is ``completely characterized by the attributes of the system regarded as a classical amplifier, and arises out of those classical amplification factors and the commutation relations of quantum mechanics.'' \cite{unruh2011}. A first necessary step, as shown in this work, therefore is the realisation of said classical amplifiers - this work provides one possible technical solution and a route towards future experiments aimed at measuring vacuum amplification from rotation and related quantum friction effects.

\section{Acknowledgements}
We acknowledge technical support from D. Grimsey and O. Warner. 
The authors acknowledge financial support from UKRI EPSRC (EP/W007444/1, EP/V000624/1 and EP/X009491/1), the Leverhulme Trust (RPG-2022-57), the QuantERA II Programme (project LEMAQUME), Grant Agreement No 101017733, and the EU Horizon Europe EIC Pathfinder project QuCoM (101046973).

\newpage
\section{Methods}

\subsection{Experimental details}\label{s:ExpDetails}

The setup is depicted in Fig.~\ref{fig:StatorField}b. 
Our Zel'dovich cylinder is a 40~mm diameter solid aluminium conductive cylinder attached to a DC motor (Maxon ECXSP19L 2 pole brushless DC motor, 0-900Hz rotation rate, as in our previous work\cite{braidottiAmplification2024}), which provides a controllable rotation speed. 
This rotating absorber is surrounded by the stator of a three-phase quadrupole induction motor (Panasonic M8MX25G4YGA). The air gap between the stator and cylinder is $\approx$1~mm. The stator consists of three sets of four coils. Each set of four are wound in alternate directions between adjacent coils in order to create a quadrupole magnetic field when a current flows.  
If the sets are provided the same alternating current with a phase difference of 120$^\circ$, the total quadrupole field rotates. As visualised in Fig~\ref{fig:ZelSimulation}a, the field rotates at half the frequency of the AC, as when the current has progressed by 180$^\circ$ the field has only rotated by a quarter circle (90$^\circ$). The rotating quadrupole field has the topological charge $m=2$. A Hall probe was used to confirm the rotation direction for a given phase ordering of the coil sets.

\begin{table*}
    \centering
    \begin{tabular}{ccccccc}
         Circuit&   C (nF)&$V_{i}$ (mV, rms)& $R_0$ at 20$^\circ$ C (Ohms)&  $R_{var}^\text{HR}$ (Ohms)&  $R_{var}^\text{LR}$ (Ohms)& Phase (rad)\\
         P1&   149.9&12.69&71.6&  22.4&  1.2&0.013\\
         P2&   149.7&12.70&71.4&  27&  1.0&-2.0915\\
 P3&  149.7&12.67& 71.4& 23.7& 1.1&2.0975\\
    \end{tabular}
    \caption{Input and component values for all three circuits. $R_0$ is given at 20$^\circ$C, but the coils heat up during the experiment, increasing the resistance (details in Supplementary). The temperature coefficient of resistance $\alpha$ for the copper wires is $\approx$ 0.004/$^\circ$C. }
    \label{tab:VoltResValues150}
\end{table*}

Each set of coils was connected in series with a capacitor and resistors, to form a resonant RLC circuit for each phase. The combination of a capacitor (which stores energy in an electric field) and inductive coils (storing energy in a magnetic field) allows energy to oscillate between them with a resonant frequency of 
\begin{equation}\label{e:resF}
    \omega_{res} = 2\pi f_{res} = \frac{1}{\sqrt{LC}}.
\end{equation}
While in the main paper frequencies $f$ and $F$ are used, in the methods section the angular frequencies $\omega = 2\pi f$ and $\Omega = 2\pi F$ will be used for simpler expressions in the mathematical theory.

\begin{figure}
    \centering
    \includegraphics[width=\linewidth]{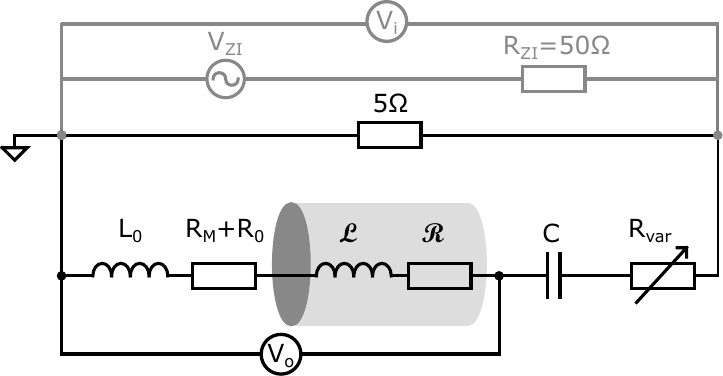}
    \caption{Circuit for each phase. When the ZI is connected as input signal $R_i = \frac{(50 \times 5)}{50 + 5}\Omega = 4.54\Omega$, otherwise $R_i = 5\Omega$  } 
    \label{fig:Circuit}
\end{figure}

The total resistance in the circuit was controlled by a fixed 5 ohm resistor and a variable resistor $R_{var}$. 
Each phase circuit input voltage was provided by one output of a Zurich Instruments (ZI) lock-in amplifier (HF2LI), which has its own internal resistance of $\approx50$~Ohm. Connecting the ZI output across the 5~ohm resistor, it acts as a voltage divider producing an effective source input voltage $V_i$, and lowered effective source impedance $R_i =$ 4.54 ohm (Fig.~\ref{fig:Circuit}). The three-gang variable resistor was used to vary the total resistance in the circuits simultaneously. This controlled whether the system was low enough resistance for the cylinder rotation to take the circuit into a total negative resistance instability regime, without changing other elements in the circuits.  

\subsection{Circuit model}

The effective RLC circuit for each phase as described in the previous section is represented in Fig.~\ref{fig:Circuit}, and includes the stator coils ($L_0$, $R_0, R_M$), the field interacting with the cylinder ($\mathcal{L}$, $\mathcal{R}$) in series with a capacitor ($C$), a fixed $5$~Ohm resistor and a variable resistor $R_{var}$. Values for each phase are given in Table ~\ref{tab:VoltResValues150}.

The circuit can be modelled in a simple way by a transfer function for complex output voltage $V_o$ (measured over the coils) from complex input voltage $V_i$ (applied over the 5~ohm resistor): 
\begin{equation}\label{eq:TransferF}
V_o = \frac{Z_{coil+cyl}}{R_i + R_{var}+Z_C+ Z_{coil+cyl} } V_i.
\end{equation}
The capacitor impedance $Z_C=1/\left( i \omega C \right)$, and the impedance of the coils and their interaction with the cylinder $Z_{coil+cyl}=Z_{cc}= R_{cc} + i\omega L_{cc}$, where 
\begin{align}
    R_{cc} &= R_0 + R_M(\omega) + \mathcal{R}(\omega,\Omega)\\
    L_{cc} &= L_0(\omega) + \mathcal{L}(\omega,\Omega).
\end{align} 
Here $R_0$ is the Ohmic resistance of the stator coil, $R_M$ accounts for any other effective resistance  present without the cylinder, such as mutual resistance between circuits or from induced currents in surroundings. $L_0$ is the inductance of the coil without the cylinder, $\mathcal{R}$ and $\mathcal{L}$ are the effective resistance and inductance contributed by the presence of the cylinder and its rotation, which is where the Zel'dovich effect appears. 

Rearranging Eq.~(\ref{eq:TransferF}) , $Z_{cc}$ can be extracted from our measurements of output voltage and known circuit values: 
\begin{equation}\label{e:Zcoilcyl}
    Z_{cc} = \frac{(Z_C + R_{circ}) V_{o}}{V_{i} - V_{o}}.
\end{equation}
The total resistance and inductance in the circuit can be found as 
\begin{align}\label{e:RLfromdata}
    R &= \Re[Z_{cc}] + R_i + R_{var}\\
    L &= L_{cc} = \Im[Z_{cc}/\omega]
\end{align}
($R$ shown in Fig.~\ref{fig:ZelSpeeds}b), and $L$ shown in supplementary materials). When the cylinder is not present ($\mathcal{R},\mathcal{L} = 0$), the total resistance of the circuit by itself is $R_{circ} = R_0+R_M(\omega)+R_i +R_{var}$, and the inductance of the circuit itself is $L_0$. By comparing the data taken with and without the cylinder present, the extra contributions of the rotating cylinder ($\mathcal{R} = R-R_{circ}$ and $\mathcal{L} = L-L_0$) are evident. 

When the circuits are exponentially amplifying, there is no measurable input signal provided, it is initially seeded by random noise, and Eq.~\ref{eq:TransferF} is no longer useful. Instead, the voltage amplitude envelope function over time $t$ is modelled by:
\begin{equation}\label{e:Exponential}
    V \propto e^{-\frac{Rt}{2L}}.
\end{equation}

\subsection{Zel'dovich cylinder model}

Firstly, from the point of view of the cylinder, a varying magnetic field impinging on a conductor induces eddy current loops in the conductor which causes a responsive magnetisation of the cylinder (a reflected field) as well as dissipation of the field energy.  This response of a material to an applied magnetic field is described by its complex magnetic susceptibility $\chi=\chi'-i\chi''$. In our case of a conductive aluminium cylinder, the dependence of the susceptibility on field frequency $\omega_-$ can be understood through the physics of penetration depth, $\delta$.

\begin{equation}\label{e:penDep}
    \delta =  \frac{1}{\sqrt{\sigma\mu\omega_-}},
\end{equation}
where $\sigma $ is the electrical conductivity of aluminium ($3.77\times 10^7 $ at 20$^\circ$C),  $\mu =  \mu_0\mu_r$ where $\mu_0$ is the vacuum permeability and $\mu_r = 1.000022$ for the relative permeability of aluminium. The penetration depth scales with frequency as $1/\sqrt{\omega}$. An EM wavelength much smaller than the conducting cylinder dimensions cannot penetrate very far, and interaction with the entire cylinder is restricted to a thin layer of eddy currents on the surface; at this extreme the reflection is strong and the absorption is weak. An EM wavelength that is very large compared to the cylinder can pass through the entire cylinder without much interaction, so both responses are weak at this extreme - like the cylinder is not present at all. When the wavelength is on the same order as the dimensions of the cylinder, the field penetrates far into the conductive cylinder without simply passing through, maximising the absorption response. 

When the cylinder is rotating at $\Omega$, an external field of frequency $\omega$ with angular momentum $m$ (with respect to the rotation axis) appears to have a rotationally Doppler-shifted frequency 
\begin{equation}\label{e:w_minus}
    \omega_- = \omega - m \Omega  
\end{equation}
in the rotating frame. One can imagine from Fig.~\ref{fig:ZelSimulation}a  how a point on the cylinder surface, rotating around the centre, will experience a different field frequency due to the way it rotates and sweeps through the spatially and temporally varying field. The cylinder's response function to a field with angular momentum changes correspondingly with rotation thanks to that shifted frequency changing the effective penetration depth $\delta(\omega_-)$.  

If the cylinder is co-rotating fast enough, $\omega_-$ can become negative. What happens to the response in that case? Zel'dovich argues{\cite{zeldovichGeneration1971,zeldovichAmplification1972,zeldovichRotating1986}} that to avoid breaking the second law of thermodynamics when conserving energy and angular momentum, when  $\omega_-$ becomes negative, the absorption (proportional to $\chi''$) also must flip sign, turning into amplification. This results in an odd-symmetric absorption response about  $\omega_-=0$.

Looking at the amplification from the point of view of the RLC circuit, the eddy currents induced in the conductive cylinder will couple a magnetic flux $\Phi^{\textrm{refl}}$ into the circuit. 
We can write the reflected flux as $\Phi^{\textrm{refl}} =\alpha(\omega_-) I$, where $I$ is the current applied to the circuit and the complex flux response function is $\alpha = \alpha'-i\alpha''$. The response function $\alpha$ has to be evaluated at the field frequency $\omega_-$ seen by the cylinder in the rotating frame. However the magnetic flux reflected into the circuit will oscillate in the laboratory frame at the applied frequency $\omega$, so the associated voltage induced in the circuit is: 
\begin{equation}
V = i\omega \Phi^{\textrm{refl}} = i\omega \alpha\left( 
 \omega_- \right) I
\end{equation}

From the complex flux response function $\alpha$, we can define the resistance $\mathcal{R}$ and the inductance $\mathcal{L}$ generated by the presence of the cylinder as:
\begin{align}
\label{m_yZ}
\mathcal{R} = Re[V/I] = \omega \alpha''\left( 
 \omega_- \right) \\
\mathcal{L} = Re[\Phi/I] = \alpha'\left( 
 \omega_- \right).
\end{align}
 
In order to find analytical expression of $\alpha$ we note that from the definitions of circuit inductance $L$ and of $\alpha$: 
\begin{equation}  \label{flux}
  \frac{\alpha}{L} = \frac{\Phi^{\textrm{refl}}}{\Phi^{\textrm{appl}}} \simeq \frac{B_r^{\textrm{refl}}}{B_r^{\textrm{appl}}}. 
\end{equation}
Here $\Phi^{\textrm{appl}} = L I$ is the self-magnetic flux applied by the current $I$. The fields $B_r^{\textrm{refl}}$ and $B_r^{\textrm{appl}}$ are the radial components of the reflected and applied fields evaluated at the coil surface. We assume that the coils are shaped to be parallel (and very close) to the cylinder external surface and neglect border effects from the finite length of the cylinder, so the fields can be assumed to be uniform across the coil area.

Now, the problem is how to calculate $B_r^{\textrm{refl}}/B_r^{\textrm{appl}}$ which requires solving the Maxwell equations for the cylinder with proper boundary conditions in the frame rotating with the cylinder. 
We use an analytical infinite cylinder model, generalizing the result derived in \cite{Perry1978, braidottiAmplification2024} for a dipolar field ($m=1$) to fields with arbitrary angular momentum $m$.
A rotating field with cylindrical symmetry of order $m$ features radial and azimuthal components rotating at $\omega_-$ with an angular dependence $\exp(-i m \varphi)$. In empty space the solutions are of the form $f(r) e^{-i m \varphi} e^{-i\omega_- t} $ with radial functions $f(r) \propto r^{m-1}$ or $f(r) \propto r^{-m-1}$. The first solution can be interpreted as the field applied by the coils (vanishing at $r=0$ for $m>1$), the second solution as the field reflected by the cylinder (vanishing at infinity). Inside the metallic cylinder the Maxwell equations take the form of the Bullard equation and the radial part can be expressed in terms of Bessel functions. Using these solutions and applying the boundary conditions at the cylinder surface, after defining:
\begin{equation}
S=\frac{B_r^{\textrm{refl}}}{B_r^{\textrm{appl}}},
\end{equation}
we find:
\begin{equation}
S = \left(\frac{a}{r}\right)^{2m} \frac{(\mu_r+1) J_m\left(\sqrt{i}\frac{a}{\delta}\right) - \sqrt{\frac{i}{m^2}}\frac{a}{\delta} J_{m-1}\left(\sqrt{i}\frac{a}{\delta}\right) }{(\mu_r-1) J_m\left(\sqrt{i}\frac{a}{\delta}\right) + \sqrt{\frac{i}{m^2}}\frac{a}{\delta} J_{m-1}\left(\sqrt{i}\frac{a}{\delta}\right)}.
\end{equation}
Here $\delta$ is the effective penetration depth in the rotating frame $\delta = 1/\sqrt{\sigma\mu (\omega-m\Omega)}$. 
Note that in the latter equation $r$ is the radius at which the coils are located, while $a$ is the radius of the cylinder. In our setup $m=2$, $a=0.020$ m and $r = 0.021$~m, so $(a/r)^{2m} \approx 0.8$.

By plugging $S$ in Eq. (\ref{flux}) we find $\alpha$, and by means of Eq. (\ref{m_yZ}) we find: 
\begin{eqnarray}\label{e:cylRL}
\mathcal{R} &=& \omega \alpha''\left(\omega_- \right) = A \, \omega L \, \text{Im}\left[S \right],\\
\mathcal{L} &=& \alpha'\left(\omega_- \right)  = A \, L \, \text{Re}\left[S \right],
\end{eqnarray} 
where $A$ is a constant of the order of 1 that can be used to take into account inaccuracies and border effects. We have used Eq. (\ref{e:cylRL}) for the theoretical curves in Fig.~\ref{fig:ZelSimulation}b and the fitting curves in Fig.~\ref{fig:ZelSpeeds}.

When the cylinder is co-rotating faster than the rotating magnetic field,  $\omega_-$ becomes negative which changes the sign of the imaginary part of $S$, 
changing the effective resistance of the cylinder $\mathcal{R}$ from positive to negative. This counteracts the positive resistance of the coil and the rest of the circuit, and if the effect is sufficiently strong, can cause the total resistance of the whole system to become negative, allowing an exponential amplification instability. 

\begin{figure}
    \centering
    \includegraphics[width=\linewidth]{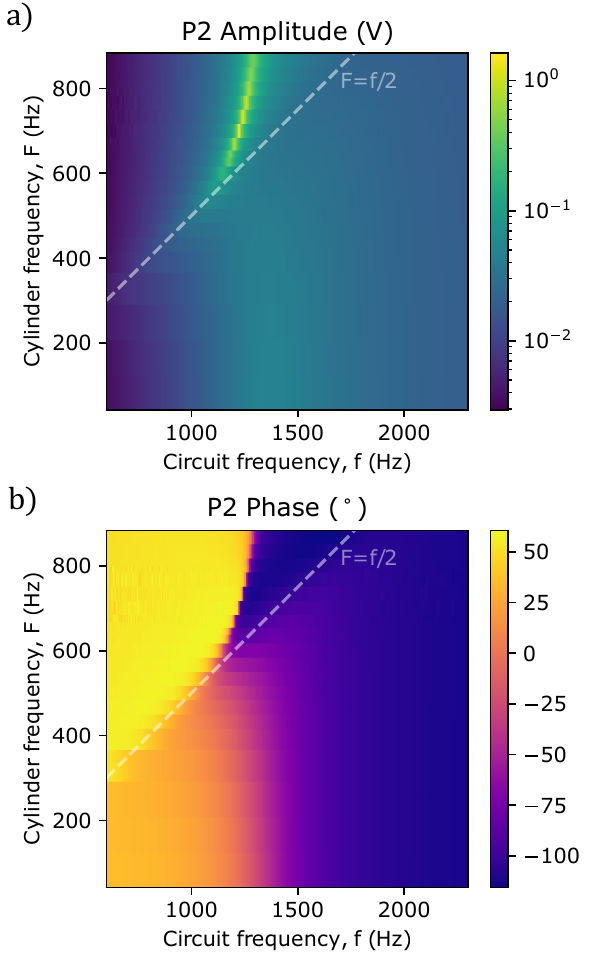}    
    \caption{The high resistance P2 amplitude (a) and phase (b) dataset, for different frequencies and rotation speeds. The Zel'dovich amplification threshold $F = f/2$ is indicated with a dashed line. The resonance peak changes in both frequency and amplitude with rotation speed. }
    \label{fig:2DAmpPhase_HR}
\end{figure}

\begin{table*}[!htp]
    \centering
    \begin{tabular}{ccccccc}
         Circuit&  $v_1$&$v_2$ (s)& $v_3$&$p_1(^\circ)$& $p_2$ ($^\circ$s)&  $p_3$ ($^\circ$s$^2$)\\
         P1&   10.82 & -\num{4.039e-4}&{0.9922} &0.5954 & -\num{5.53e-3} & \num{1.507e-6}\\
         P2&  11.23 &-  \num{5.069e-4} & {0.9955} &-4.186&  0&  0\\
 P3&  11.42 &- \num{5.837e-4} & {0.9971}&3.788& -{0.01007}& \num{2.494e-6}\\
    \end{tabular}
    \caption{Values for attenuated measurement conversion corrections.}
    \label{tab:conversioncoeffs}
\end{table*}

\subsection{Experimental procedure: Stable regime}

For the stable regime the ZI HF2LI was used to supply the input signal ($V_i$ at $f$) and measure the rms (root-mean-square) output signal ($V_o$, at the input frequency $f$) amplitude and phase with respect to the input reference.  
As the HF2LI models only have two outputs and two inputs each, two were used, one locked into the other, to always provide the three excitation voltages $V_i$ at the same frequency with a fixed 120$^\circ$ phase difference between them. Only the HF2LI providing the initial oscillator reference could perform lock-in measurements of $V_o$ at a reasonable speed, so while all three phases were always energised with $V_i$ while taking data, only two phases $V_o$ were measured at once. To get data from all phases, runs were repeated, swapping out one of the measured phases. 

The dependency of the steady-state voltage output on both circuit frequency ($f$) and cylinder rotation frequency ($F$) was investigated. The cylinder frequency was set and strictly maintained with closed-loop control by the ESCON controller for the maxon DC motor. The direction of cylinder rotation could also be chosen. For a set cylinder rotation speed (in the range 0 to 900 Hz), and also for the case of no cylinder present (to determine $R_{circ}$ and $L_0$), voltage measurements were taken while the circuit frequency was swept, usually recording at 250 points from 600 - 2600 Hz. Fig.~\ref{fig:2DAmpPhase_HR} shows the measured $V_o$ amplitude and phase for the three circuits as a function of $f$ and $F$ when $R_{var}$ has its higher value. This data was then analysed using Eqs.~\ref{e:Zcoilcyl} and \ref{e:RLfromdata} to extract $R$ and $L$ measurements.

\subsubsection{Attenuated measurements}
When the amplification was high, the resonance peaks would saturate the ZI's 1V (rms) input range, which could lead to an underestimate of the output voltage. To avoid this, data at some rotation frequencies (667 to 800Hz) were taken with the measurement probes on 10X attenuation mode. However it came with the issue that the recorded voltage was not a exact 0.1 multiple of the 1X recorded voltage, but the amplitude and phase depended on magnitude and frequency. Small corrections to the conversion were needed to incorporate these few data points with the normal 1X measurements dataset. 
 
Data taken at the same cylinder rotation (667~Hz and 800~Hz) in both 1X and 10X modes was used to fit modified conversion functions for our amplitude and phase data:
\begin{align}\label{e:AttConv}
    |V| &= (v_1 + v_2 f ) |V_\text{10X}|^{v_3} \\
    \phi &= \phi_\text{10X} + p_1 + p_2 f + p_3 f^2 
\end{align}
with values for each measurement channel shown in Table~\ref{tab:conversioncoeffs}. The noise level for those measurements is also amplified, which becomes noticeable in the derived values at the tails of the resonance peaks where the amplitude is very low (Fig.~\ref{fig:2DR_andpredR} in the supplementary material for example). 

\subsection{Experimental procedure: unstable regime}

To investigate the unstable exponential amplification regime, no input signal was required, so the input and outputs of the ZI were removed, and the three phases were measured simultaneously with an oscilloscope (Tektronix DPO2024B). When this voltage was measured over the coils, the oscilloscope would saturate (Fig.~\ref{fig:otherselfosc} in Supplementary materials), whereas measuring over the 5 Ohm resistor allowed the peak voltages at the turning point to be recorded fully (Fig.~\ref{fig:self-osc}). The cylinder speed would be set by the DC motor control to be in the exponential amplification region. Here an open loop control\cite{ESCON2024} was used to set a demand speed the system would try to achieve by sending a voltage proportional to the speed and the drawn motor current. This is a slower and less strict adjustment than the closed loop control which uses the motor's actual speed for feedback. 
The open-loop control allowed the motor speed to drop below the instability threshold when the cylinder was transferring large amounts of energy to the RLC circuit, which cut off the process, this limiting mechanism saving the RLC circuits from getting fried repeatedly under the more strict maintenance of the closed-loop control.

From the oscilloscope data, the spectrogram is generated directly from the data, with the size of the FFT block chosen to produce a spectrogram with reasonable trade off between time and frequency resolution. For other measurements a Hilbert transform is used to extract the instantaneous amplitude, phase and frequency of the signal. This extracts the peak voltage amplitude, not the rms ($V_\text{peak} = V_\text{rms}\sqrt{2}$). An optional butterworth bandpass filter can be used to filter away the noise outside the frequency region of the signal. The data is smoothed by averaging over time (e.g. from 12.5 kHz sample rate to 62.5Hz). The regions of each channel, where it measures within its measurement range, are combined to make a single voltage envelope signal from the three channels. $L$ is calculated from the frequency using Eq.~\ref{e:resF}, then combined with the instantaneous amplitude exponent value to calculate $R$ from Eq.~\ref{e:Exponential}.

\bibliography{3PhaseZeldyPaper1213}


\newpage
\section{Supplementary}
\subsection{Temperature}

The coils would heat up when the cylinder was rotating at high speeds, this did change their resistance accordingly. 
\begin{equation}
    dR = \alpha  dT R_{ref}
\end{equation}
\begin{equation}\label{e:tempRchange}
    R = R_{ref}(1 + \alpha(T-T_{ref})) 
\end{equation}

\begin{figure}
    \centering
    \includegraphics[width=\linewidth]{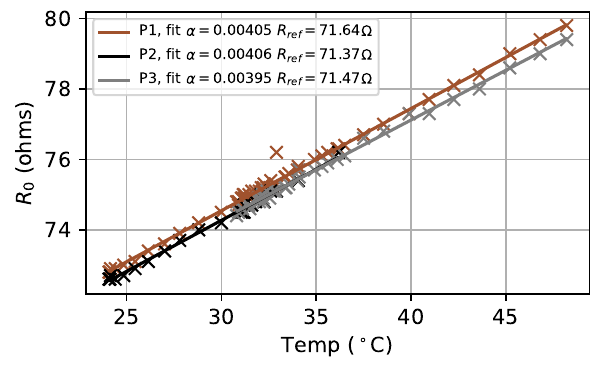}
    \caption{Coil casing temperature and coil resistance data taken over several different measurement runs. A fit to Eq.~\ref{e:tempRchange} for $T_{ref}=20^\circ C$ is given.}
    \label{fig:tempalphas}
\end{figure}

The standard temperature coefficient of resistance $\alpha$ for copper at $T_{ref}=$20$^\circ$C is around 0.004/$^\circ$C.
The temperature of the stator casing was measured by a thermocouple probe and the ohmic resistance of the coils taken with a multimeter during measurements, these values agreed with the standard alpha value (Fig.~\ref{fig:tempalphas}). 

\begin{figure}
    \centering
    \includegraphics[width=\linewidth]{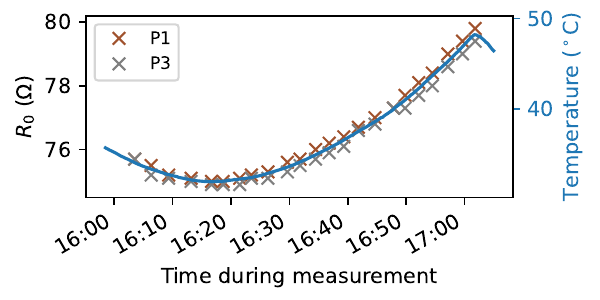}
    \caption{Coil resistance and stator temperature over time during a long measurement run, during which the cylinder speed is stepped up from 0 Hz to 867~Hz.}
    \label{fig:tempovertime}
\end{figure}

The temperature change over a dataset would depend on how long the dataset took, i.e. the number of cylinder speeds sampled. During the longest datasets, the high resistance and positive direction runs in which the cylinder speed was increased in steps between 0~Hz and 900~Hz over the duration of around 50 minutes, the temperature of the stator casing could increase by 10-20$^\circ$C. The coils were not usually left to cool to room temperature between runs, so the temperature would also cool during the slower cylinder rotation speeds of the runs (Fig.~\ref{fig:tempovertime}). Even on the longest runs, the resistance did not usually vary by more than 5 ohms.
 
As this change in resistance is small compared to the change generated by the cylinder rotation rate, for simplicity it has not been considered in the main data analysis, where the offset resistance $R_{circ}$ is taken to be constant over the cylinder rotation $F$. A temperature change would also change the conductivity $\sigma$ of the cylinder, again this effect is small enough to be neglected, and the conductivity is considered constant in the Zel'dovich model. 
In the exponential amplification regime, the measurements are taking place over the order of seconds, so the temperature can be considered constant over the measurement.

\subsubsection{Inter-circuit interactions}

The circuit model (Eq.~\ref{eq:TransferF}) generally considers each phase independently, but incorporates the inter-circuit interactions from the shared magnetic field in a limited way, through the $R_M(\omega)$
and $L_0(\omega)$ parameters for each phase, which make up part of $Z_{cc}$.

\begin{figure}
    \centering
    \includegraphics[width=\linewidth]{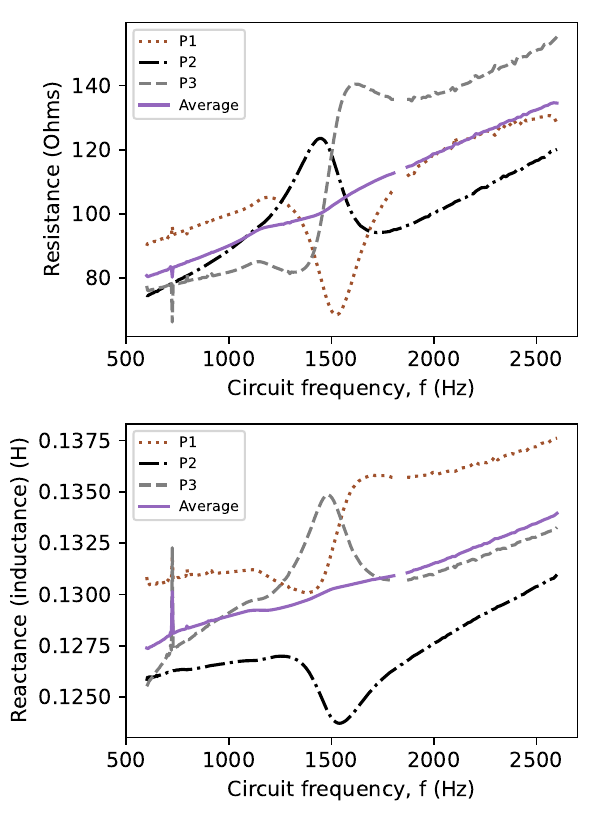}
    \caption{No cylinder data (high resistance case). Shows resistance $R_0 + R_M$ and inductance $L_0$ extracted from measurements over coils.}
    \label{fig:NoCylRL}
\end{figure}

The need for the frequency dependence of these parameters is apparent when $Z_{cc}$ is measured without the cylinder present ($\mathcal{R}$, $\mathcal{L} = 0$), see Fig.~\ref{fig:NoCylRL}. We see features for each phase centred near $f = 1500~$Hz. When the contributions of all three phases are summed together, the features cancel out into a simpler, approximately linear frequency dependence over the measured range, indicating an origin from some shared interactions between the circuits, likely due to the fact they are not exact copies.

A full treatment of the system would require a matrix approach rather than a circuit-by-circuit approach, to fully take into account the inter-circuit interaction. Nevertheless, the simple model we have used is more than adequate to reveal the presence of the Zel'dovich effect in this experiment. Furthermore, the majority of this contribution is circuit frequency dependent rather than cylinder rotation speed dependent, and so when slicing the dataset to analyse results at a constant circuit frequency (Fig.~\ref{fig:ZelSpeeds}) it can be approximated simply as a background offset for each phase, absorbed into $R_{circ}$ and $L_0$.

\begin{figure}
    \centering
    \includegraphics[width=\linewidth]{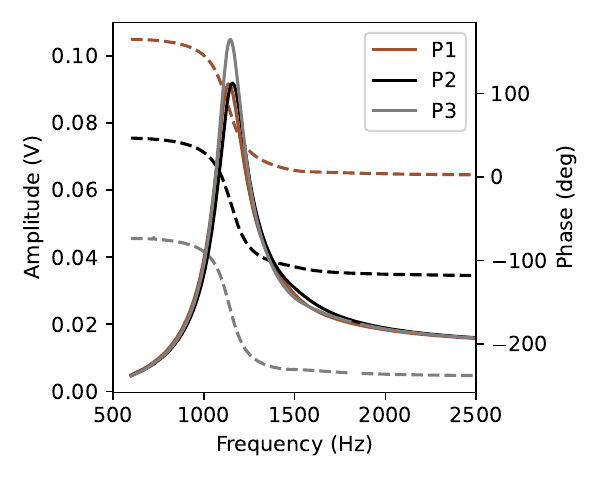}    
    \caption{No cylinder measurement data from all three phases, high resistance case. Shows the output r.m.s.\ voltage $V_o$ amplitude (solid line) and phase (dashed line) resonance over the circuit frequency range. } 
    \label{fig:ZelNoCyl}
\end{figure}

\begin{figure}
    \centering
    \includegraphics[width=\linewidth]{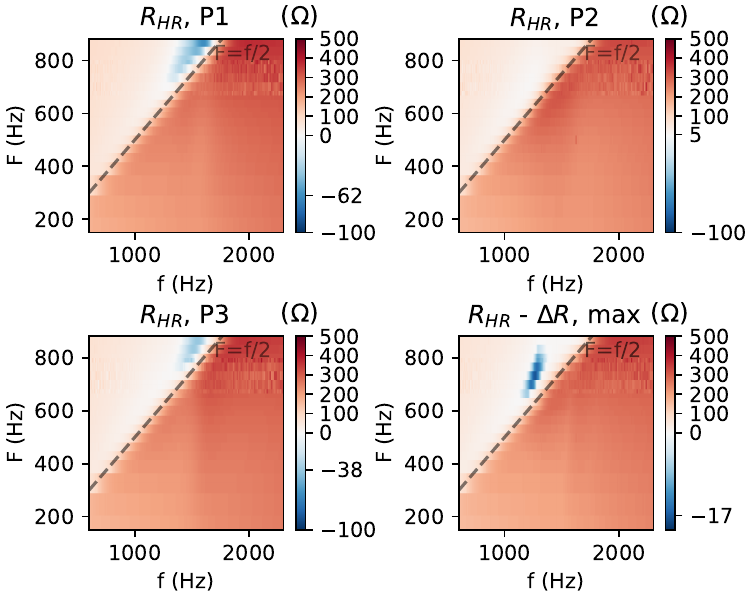}
    \caption{The high resistance datasets, with the Zel'dovich amplification threshold $F = f/2$ indicated. Extracted total resistance $R$, all three phases. Also in the bottom right, a prediction of where, for the low-resistance case, the system has for all circuits a negative resistance. The high resistance extracted total resistance $R$ data for all three circuits is offset by the respective changes in $R_{var}$, and the maximum value of the three circuits is shown. The negative resistance lies along the resonance peak curve in \ref{fig:2DAmpPhase_HR}. Note the two scales for positive and negative resistance.}
    \label{fig:2DR_andpredR}
\end{figure}

\subsection{Estimate of instability region}

Fig.~\ref{fig:ZelSpeeds}b in the main paper shows the measured total circuit resistance at $f=1181$~Hz for all three circuits indicates a range of $F$ where, when $R_{var}$ is low, all the circuits are expected to have a total negative resistance to a $f=1181$~Hz signal. However, as this is only for one value of $f$, the range of $F$ where the system is unstable to any $f$ will be greater. 
Fig.~\ref{fig:2DR_andpredR} shows how the  total resistance varies with both $f$ and $F$ in the $R_{var}$ high case. Note that while at some $f,F$, some circuits individually have a total negative resistance, in this parameter region there is always one circuit (P2) that is still positive in this $R_{var}$ high regime, and thus the overall coupled system is still stable, and is able to be measured without any exponential increase in the voltage (or things blowing up) over time.
The bottom right plot in Fig.~\ref{fig:2DR_andpredR} shows (akin to the grey offset in Fig.~\ref{fig:ZelSpeeds}b) an expectation of where, in the low resistance case, the system would be unstable - where all three circuits are expected to have a total negative resistance. This prediction is made from offsetting the high resistance data by $\Delta R$ for each circuit, and then plotting the maximum value out of the three circuits (note this may not always be P2, due to the circuit interactions mentioned in the previous section, shown in Fig.~\ref{fig:NoCylRL}). The region where there is a negative resistance predicted follows the peak amplitude region e.g.\ in Fig.~\ref{fig:2DAmpPhase_HR}. 

\subsection{Measured inductance}

\begin{figure}
    \centering
    \includegraphics[width=\linewidth]{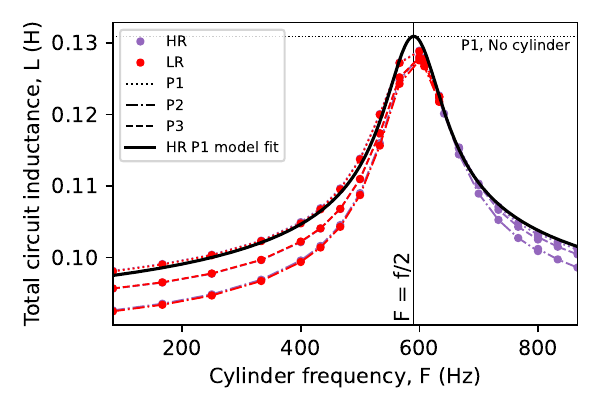}
    \caption{Total inductance in the three circuits P1, P2, P3 at $f=1181$~Hz for different cylinder rotation frequencies $F$. Purple dots are data for the high resistance measurements, and coinciding with these as expected, the red dots are the low resistance measurements. Various dashed lines are the linear interpolations between datapoints. The vertical line indicates the Zel'dovich threshold rotation, the dotted horizontal line the no-cylinder measured P1 value (0.131~H) in the high-resistance case. The Zel'dovich model fits to the P1 high resistance data are also plotted on top (black solid line), with the coupling strength $A=0.397$ and a constant $L_0=0.131$ H being free fit parameters.}
    \label{fig:ExpInductancesALL}
\end{figure}

While the real part of the voltage measurements is used to calculate the resistance $R$ in Fig.~\ref{fig:ZelSpeeds}b of the main paper, the imaginary part can be used to calculate the inductance $L$, shown in Fig.~\ref{fig:ExpInductancesALL}. Included is a fit of the inductance change in the P1 data to that predicted by the Zel'dovich theory. Here the same $A$ value is used from the fit in Fig.~\ref{fig:ZelSpeeds}b, and the $L_0$ offset is used as a free parameter, with the fitted value $L_0=0.131$ H matching the measured P1 no-cylinder value at that circuit frequency $f$, and the data following the model curve well.

\begin{figure}
    \centering
    \includegraphics[width=\linewidth]{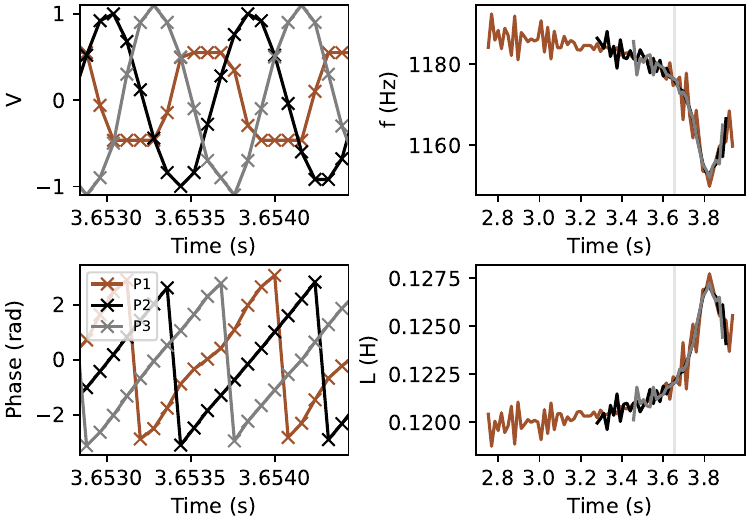}
    \caption{Signal, phase, instantaneous frequency, calculated inductance for the demand speed $F=+643$~Hz dataset shown in Fig.~\ref{fig:self-osc}. The signal and phase are shown for a short time slice, the position of which is indicated by a vertical line in the $f$ and $L$ plots. The phase relationship between the circuits shows the excited EM mode has a definite rotation direction that is co-rotating with the cylinder.}
    \label{fig:selfoscSMadditions}
\end{figure}

 \begin{figure*}
     \centering 
     \includegraphics[width=\linewidth]{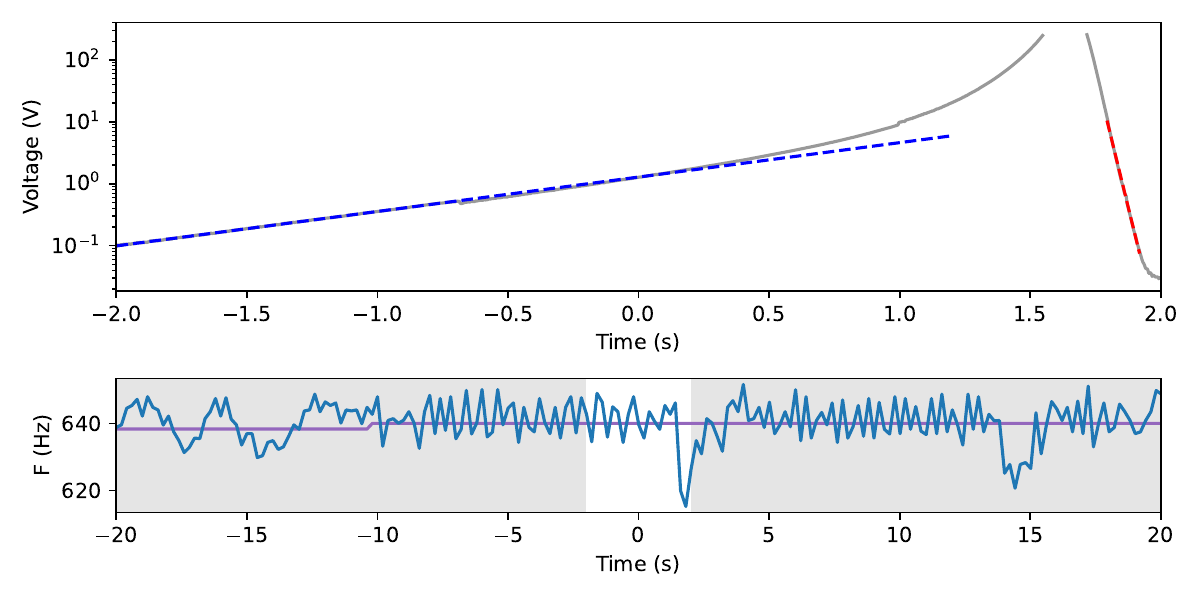}
     \caption{Other self oscillation data. 640~Hz cylinder rotation. Above shows the voltage measured over the coil. Here fits are done using Eq.~\ref{e:Exponential} to regions where there is simple exponential behaviour to the amplification or the decay. Initial slope up (blue): 1.28 $e^{1.3 t}$, slope down (red):(2.94e-03) $e^{-39.60 (t-2)}$.
     Below is shown the recorded motor speed (blue), and the speed setting (purple). At around -10s the set speed is increased. The region with the white background corresponds to the time period of the voltage measurement above.}
     \label{fig:otherselfosc}
 \end{figure*}

\begin{figure}
    \centering
    \includegraphics[width=\linewidth]{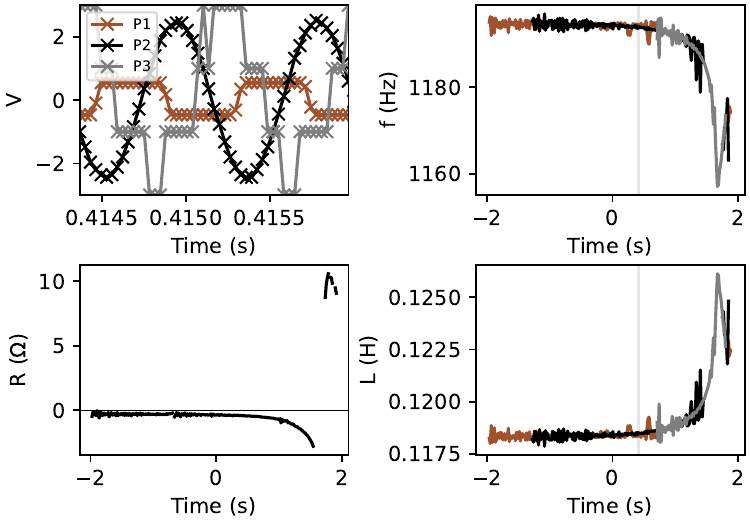}
    \caption{Voltage signal, instantaneous frequency, calculated resistance and calculated inductance for the demand speed $F=+640$~Hz dataset shown in Fig.~\ref{fig:otherselfosc}. The signal is shown for a short time slice, the position of which is indicated by a vertical line in the $f$ and $L$ plots.}
    \label{fig:otherselfoscSMadditions}
\end{figure}

\begin{figure}
    \centering
    \includegraphics[width=\linewidth]{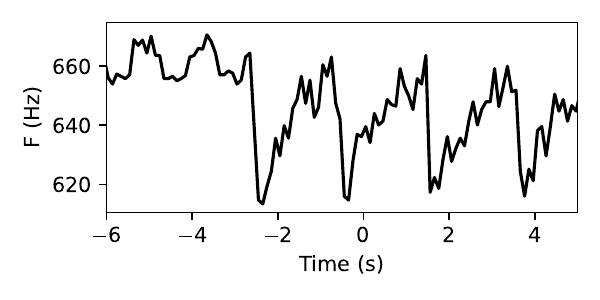}
    \caption{Cylinder speed data from the motor corresponding to the voltage measurement shown in the inset of Fig.~\ref{fig:self-osc}a. }
    \label{fig:660hz_speeds}
\end{figure}

\subsection{Additional exponential amplification data}

Fig.~\ref{fig:selfoscSMadditions} shows the voltage signal, phase, instantaneous frequency and $L$ values from the Fig.~\ref{fig:self-osc} $F=$+643~Hz driven data. Each circuit is measured with a different oscilloscope range to together track the exponentially increasing signal over more orders of magnitude. This is why for the time slice chosen, the voltage signal for P1 is saturated. The measured phases of the three circuits show when P1 has zero phase, P2 has -2.1 radians, P3 has +2.1 radians. This is the same 120$^\circ$ phase relationship between P1, P2 and P3 as in Table~\ref{tab:VoltResValues150} used to create the co-rotating mode for a positive rotation speed. This shows the exponentially amplifying EM mode created by the instability has a definite rotation direction, and that it is co-rotating with the cylinder, as expected from the Zel'dovich model.

Fig.~\ref{fig:otherselfosc} shows additional exponential amplification data. In this measurement, the voltage is measured over the coils, rather than the 5$\Omega$ resistor. Due to the large impedance of the coils ($Z \approx \omega L$) this measurement saturates the $\pm$300V max range of the oscilloscope in the peak super-exponential region, but allows more of the exponential region to be observed. 

The speed data is also shown for this measurement. The speed data was taken while the resistors were on their low values, while the speed is stepped up slowly from below the instability threshold to just above it. The step in the purple line shows the moment of the step in speed prior to the observed signal. A few seconds later a signal emerges over the oscilloscope noise floor and is recorded in the time period -2 to +2s. As the signal peaks and saturates the drop in motor speed is recorded, as is the exponential decay of the signal. The motor speed picks up again, and a second drop in motor speed arises from the subsequent exponentially amplified signal peak. The signals are seeded from noise in the circuit lower than the oscilloscope noise on the measurement, at a minimum this is thermal noise, the thermal noise associated with the inductive coils would be $I_\text{r.m.s.} = \sqrt{\frac{k_B T}{L}}$. 

Fig.~\ref{fig:otherselfoscSMadditions} shows the corresponding signal, instantaneous frequency, extracted resistance and inductance. For this rotation speed just within the instability region the extracted negative resistance in the exponential region is very small, around -0.3 ohms, similar to that in Fig.~\ref{fig:self-osc}d. 

To see how see how the speed of the cylinder affects the growth timescale of the amplification and the revival period of the peaks, we can compare the two measurements taken at F= +640 and +643~Hz with that taken at F=+660~Hz, the inset data of Fig.~\ref{fig:self-osc}a. 
There an initial cylinder speed of 660Hz (Fig.~\ref{fig:660hz_speeds}) is obtained before $R_{var}$ is switched to low values. The first excitation has an extracted negative resistance of -6.6 ohms in its constant-exponential region. After the first excitation drops the speed of the motor, the motor does not have time to fully get back up to the set original speed before the next growing excitation is above the oscilloscope noise floor (and not even before the self-halting mechanism kicks in). Therefore the subsequent peaks exponentially increase more slowly and so have a smaller extracted negative resistance of around -5 ohms (although as the speed is increasing in this case, there is no approximately constant exponent region).

\end{document}